\documentclass{appolb}
\usepackage{epsfig}

\begin{document}
\title{$\eta$ and $\eta$' production in nucleon-nucleon collisions\\
near thresholds%
\thanks{Presented at ''Symposium on Meson Physics'', 1 - 4 October 2008, Cracow, Poland.}%
}
\author{L. P. Kaptari
\address{Research Center Dresden-Rossendorf, 01314 Dresden, PF 510119, Germany\\
Bogoliubov Laboratory of Theoretical Physics, 141980, JINR,  Dubna, Russia}
\and
B. K\"ampfer
\address{Research Center Dresden-Rossendorf, 01314 Dresden, PF 510119, Germany\\
TU Dresden, Institut f\"ur Theoretische Physik, 01062 Dresden, Germany}
}
\maketitle
\begin{abstract}
The production of $\eta$ and $\eta$' mesons in nucleon-nucleon collisions
near thresholds is considered within a one-boson exchange model.
We show the feasibility of an experimental access to transition
formfactors.
\end{abstract}
\PACS{13.75.-n, 13.75.Cs, 14.20.Gk}

\section{Introduction}

The pseudo-scalar mesons $\eta$ and $\eta'$
represent a subject of considerable interest since some time
(cf.\ \cite{diekmann} 
for surveys).
Investigations of various aspects of $\eta$ and $\eta'$ mesons are tightly related with
several theoretical challenges and can augment the experimental
information on different phenomenological
model parameters. For instance, the "anomalously" large mass of the $\eta'$ meson,
as member of the $SU_A(3)$ nonet \cite{Bass}, can be directly connected with the $U(1)$
axial anomaly in QCD. A combined phenomenological analysis of
$\eta$ and $\eta'$ production in $N+N$ reactions
together with the $U_A(1)$ anomaly provides additional information on the
gluon-nucleon coupling, which can be used to describe, e.g., the so-called "spin crisis".
Also, the knowledge of the nucleon-nucleon-$\eta'$ coupling strength 
allows a better understanding of the origin of the OZI rule violation in $N+N$ reactions.
A remarkable fact is that near the threshold
the invariant mass of the $NN\eta'$  system
in such reactions is in the region of heavy nucleon resonances, i.e.
resonances with isospin $1/2$ can be investigated via these processes.
Furthermore, 
the so-called "missing resonances" can be studied.

Another aspect of $\eta$ and $\eta'$ production in elementary hadron reactions is
that both mesons have significant Dalitz decay channels into
$e^+ e^- \gamma$. As such, they constitute further sources of di-electrons.
It is, in particular, the $\eta$ which is a significant source of $e^+ e^-$
pairs, competing 
with $\Delta$ Dalitz
decays and bremsstrahlung, as the analysis \cite{eta_contributions}
of HADES data \cite{HADES} shows.
One of the primary aims of the HADES experiments \cite{HADES} is to seek
for signals of chiral symmetry restoration in compressed nuclear matter. For such an
endeavor one needs a good control of the background processes, including
the $\eta'$ Dalitz decay, in particular at higher beam energies,
as becoming accessible at SIS100 within the FAIR project \cite{FAIR}.

The $\eta$ and $\eta'$ Dalitz decays depend on the pseudo-scalar transition form factor, which
encodes hadronic information accessible in first-principle QCD calculations or
QCD sum rules. 
The Dalitz decay process of a pseudo-scalar meson $ps$ can be presented as
$ ps \to \gamma +\gamma^* \to \gamma  + e^- + e^+$.
Obviously, the probability of emitting a virtual photon is
governed by the dynamical electromagnetic
structure of the "dressed" transition vertex $ps \to \gamma \gamma^*$ which
is encoded in the transition form factors.
If the decaying particle
were point like, then calculations of mass distributions
and decay widths would be straightforwardly given by QED.
Deviations of the measured quantities from
the QED predictions directly reflect the
effects of the form factors and thus the internal hadron structure.

The present paper reports parameterizations of $\eta$ and $\eta'$
production cross sections
in nucleon-nucleon collisions near the respective thresholds within
a one-boson exchange model. Emphasis is put on the accessibility of
transition formfactors encoding the strong-interaction $\eta, \, \eta'$ structure.

\section{One-boson exchange model}

Cross sections of interest are \cite{our_eta,our_eta_prime}
\begin{eqnarray}
&& d^5\sigma^{tot}_{NN\to NN{ps}}=\frac{1}{2(2\pi)^5\sqrt{\lambda(s,m^2,m^2)}} \nonumber\\
&& \times  \frac14
\sum_{spins}| T_{NN\to NN{ps}}|^2 ds_{1'2'}
dR_2^{N_1N_2\to s_{ps} s_{1'2'}}
dR_2^{s_{1'2'}\to N_1'N_2'}
\end{eqnarray}
with two-particle invariant phase space $R_2^{ab\to
cd}=\sqrt{\lambda(s_{ab},m_c^2,m_d^2)}/(8s_{ab}) d\Omega^*_c$ for
the production of $ps \equiv \eta, \eta$' and

\begin{eqnarray}
\frac{d\sigma}{ds_{ps} ds_{\gamma^*}} =
\frac{d\Gamma_{{ps}\to \gamma e^+e^-}}{ds_{\gamma^*}}\,
\frac{1}{4\pi\sqrt{s_{ps}}}
\frac{1}{\left(\sqrt{s_{ps}}-m_{ps}\right)^2 +\frac14 \Gamma_{ps}^2}
d^5\sigma^{tot}_{NN\to NN{ps}}
\label{two}
\end{eqnarray}
for the Dalitz decay. Integrating the latter one over $ds_{ps}$
or taking it at $s = m_{ps}^2$ is meant to access the electromagnetic formfactors
appearing in
\begin{eqnarray}
\frac{d\Gamma_{{ps}\to \gamma e^+e^-}}{ds_{\gamma^*}} =
\frac{2\alpha_{em}}{3\pi s_{\gamma^*}} \left(
1-\frac{m_{ps}^2}{s_{\gamma^*}}\right)^3 \Gamma_{{ps}\to  \gamma
\gamma }  \left | F_{ {ps} \gamma \gamma^*\,}(s_{\gamma^*})\right
|^2. \label{dgamma}
\end{eqnarray}

\subsection{$\eta$ channel}

\begin{figure}[h]  
\vskip -1mm
\includegraphics[width=1.0\textwidth]{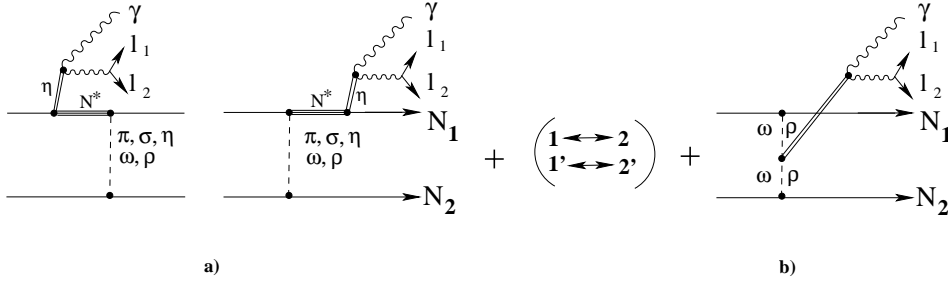} %
\vskip -3mm
\caption{Diagrams for the process
$N N \to N N  \gamma l_1 l_2$ within the one-boson exchange model.
a) Dalitz decays of $\eta$ mesons from bremsstrahlung like diagrams.
The intermediate
baryon $N^*$ (triple line) can be either a nucleon or a nucleon resonance 
($S_{11}(1535)$, $P_{11}(1440)$, $D_{13}(1520)$).
Analog diagrams for the
emission from Fermion line $N_2$.
b) Dalitz decay of $\eta$ mesons from internal meson conversion.
Exchange diagrams are not displayed.
Later on we identify $l_{1,2} = e^\pm$ and denote the di-electron invariant
mass by $s_{\gamma*}$.}
\label{fig2}
\end{figure}

We employ here a one-boson exchange model, where the $\eta$
production is described by the diagrams exhibited in Fig.~\ref{fig2}.
The sum of these diagrams generate the invariant amplitude
$T_{NN\to NN{ps}}$ via interaction Lagrangians.

\subsection{$\eta$' channel}
The calculation of $\eta$' uses the same diagram topology as in Fig.~\ref{fig2} 
(with $\eta \rightarrow  \eta'$)
supplemented by $a_0$ exchange in a). 
The included resonances are $S_{11}(1650)$ with odd parity, and
$P_{11}(1710)$ and $P_{13}(1720)$ with even parity.

\subsection{Interaction Lagrangians}

The employed interaction Lagrangians can be represented as follows.\\
(i) Nucleon currents:
\begin{eqnarray}
{\cal L}_{\sigma NN }&=& g_{\sigma NN} \bar N  N \it\Phi_\sigma ,\\
{\cal L}_{a_0 NN }&=& g_{a_0 NN} \bar N (\tau \Phi_{a_0}) N \it ,\\
{\cal L}_{\pi NN}&=&
-\frac{f_{ \pi NN}}{m_\pi}\bar N\gamma_5\gamma^\mu \partial_\mu
({\tau \Phi_\pi})N , \\
{\cal L}_{\eta NN}&=&
-\frac{f_{\eta NN}}{m_\eta}\bar N\gamma_5\gamma^\mu \partial_\mu
\Phi_\eta N ,\\
{\cal L}_{\rho NN}&=&
-g_{\rho NN }\left(\bar N \gamma_\mu{\tau}N{\Phi_ \rho}^\mu-\frac{\kappa_\rho}{2m}
\bar N\sigma_{\mu\nu}{\tau}N\partial^\nu{\Phi_\rho}^\mu\right) ,\\
{\cal L}_{\omega NN}&=&
-g_{\omega  NN }\left(
\bar N \gamma_\mu N {\it\Phi}_{\omega}^\mu-
\frac{\kappa_{\omega}}{2m}
\bar N \sigma_{\mu\nu}  N \partial^\nu \it\Phi_{\omega}^\mu\right),
\end{eqnarray}
(ii) Spin $\frac12 $ resonances ($S_{11}$ and $P_{11}$):
\begin{eqnarray}
{\cal L}_{NN^*ps}^{(\pm)}(x)&=&\mp \frac{g_{NN^* ps} }{m_{N^*}  \pm m_N}
 \bar\Psi_R(x)\left\{
\begin{array}{c}
\gamma_5\\ 1\end{array} \right\} \gamma_\mu\partial^\mu \Phi_{ps}(x) \Psi_N(x) + h.c. \\
{\cal L}_{NN^*V}^{(\pm)}(x)&=&  \frac{g_{NN^* V} }{2(m_{N^*}  + m_N) }
 \bar\Psi_R(x)\left\{
\begin{array}{c}
1\\ \gamma_5 \end{array} \right\} \sigma_{\mu\nu} V^{\mu\nu}(x) \Psi_N(x) + h.c. 
\nonumber \\[-3mm]
\end{eqnarray}
(iii) Spin $\frac32 $ resonances ($D_{13}$ and $P_{13}$):
\begin{eqnarray}
{\cal L}_{NN^*ps}^{(\pm)}(x) &=&  \frac{g_{NN^* ps} }{m_{ps}}
 \bar\Psi_R^\alpha(x)\left\{
\begin{array}{c}
1\\ \gamma_5\end{array} \right\}  \partial^\alpha \Phi_{ps}(x) \Psi_N(x) + h.c. \\
{\cal L}_{NN^*V}^{(\pm)}(x)& = & \mp i\frac{g_{NN^* V}^{(1)} }{2m_N }
 \bar\Psi_R^\alpha(x)
\left\{\begin{array}{c}\gamma_5\\ 1\end{array} \right\}
\gamma_{\lambda} V^{\lambda\alpha}(x) \Psi_N(x)    \nonumber \\
& - &
\frac{g_{NN^* V}^{(2)} }{4m_N^2 } \partial_\lambda \bar\Psi_R^\alpha(x)\left\{
\begin{array}{c}
\gamma_5\\ 1\end{array} \right\} V^{\lambda\alpha}\Psi_N(x) +h.c.
\end{eqnarray}
with the abbreviations $ps \equiv \pi$ or $\eta$ or $\eta$',
$\Phi_{ps} \equiv( \tau \Phi_\pi(x)) $ or
$\Phi_{\eta'}(x)$,
$V \equiv V_\omega(x)$ or
$V(\tau\rho(x))$, and
$V^{\alpha\beta}=\partial^\beta V^\alpha - \partial^\alpha V^\beta$.
Furthermore needed interactions, such as
${\cal L}_{ps \omega\omega}$,
${\cal L}_{ps \rho\rho}$,
${\cal L}_{\gamma ll }$, and
${\cal L}_{ps \gamma\gamma}$, are listed in \cite{our_eta_prime}.

\subsection{Formfactors}

Strong formfactor are needed to dress the nucleon -- nucleon (resonance) -- meson
vertices. These are listed in detail in \cite{our_eta,our_eta_prime}.

The electromagnetic formfactors encode non-perturbative transition matrix elements
$F_{ps \gamma \gamma*}$ in (\ref{dgamma}),
basically accessible within QCD. Here, however, we contrast a few
parameterizations: (i) so-called QED formfactor meaning a structure-less
particle with $\left | F_{ ps  \gamma \gamma^*\,}(s_{\gamma^*})\right |^2 = 1$,
(ii) a parametrization suggested by the vector meson dominance (VMD) model
\begin{equation}
F^{VMD}_{ ps \gamma \gamma^*\,}(s_{\gamma^*})
\sum_{V=\rho,\omega,\phi} C_V \frac{m_V^2}{\hat m_V^2-s_{\gamma^*}},
\label{fvmd}
\end{equation}
with $F_{ps \gamma \gamma^* }(s_{\gamma^*}=0)=1$,
$\sum_V C_V = 1$ and $\hat m_V = m_V - i \Gamma_V/2$.
The values of $C_V$ are quoted in \cite{our_eta_prime}.
For the case of $\eta$, the kinematically accessible region is restricted
and, as a consequence, the $\rho$ contribution is sufficient.
(iii) For $\eta$', a monopole fit
$ F_{\eta' \gamma \gamma^* }(Q^2) = (1-Q^2/\Lambda_{\eta'})^{-1}$
\cite{our_eta_prime} may be used,
which does not differ too much from the VMD parametrization.

\subsection{Initial state and final state interactions}

Initial state interactions are accounted for by effective
reduction factors for
${}^3 P_0$, ${}^1 P_1$ waves: $\zeta = 0.277 \, (pp), \, 0.243 \, (np, pp)$
\cite{nakayama_eta}.
Final state interactions are treated by the Jost function formalism,
see \cite{Titov} for details.

\begin{figure}[h]  
\vskip -3mm
\includegraphics[width=.5\textwidth]{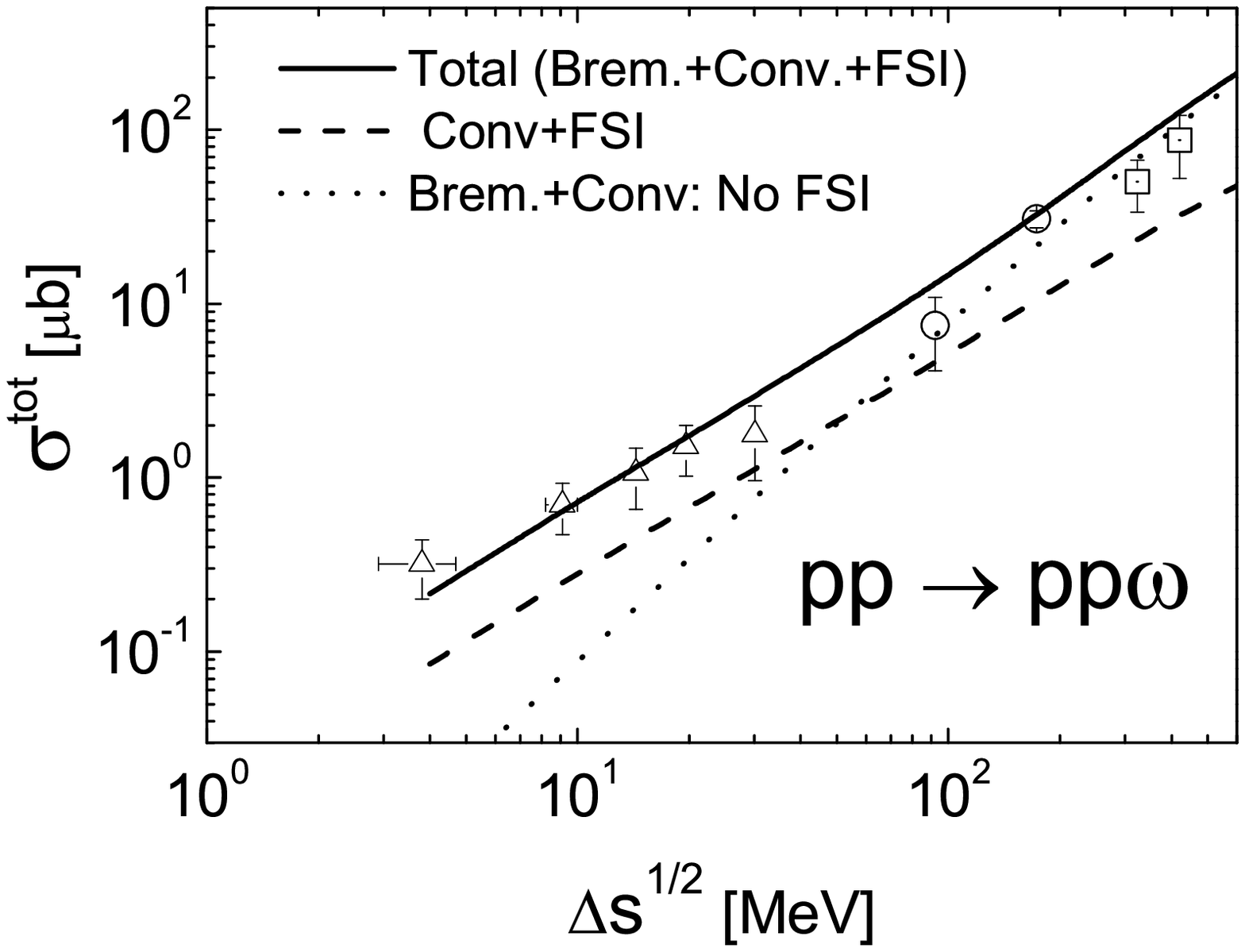} %
\includegraphics[width=.5\textwidth]{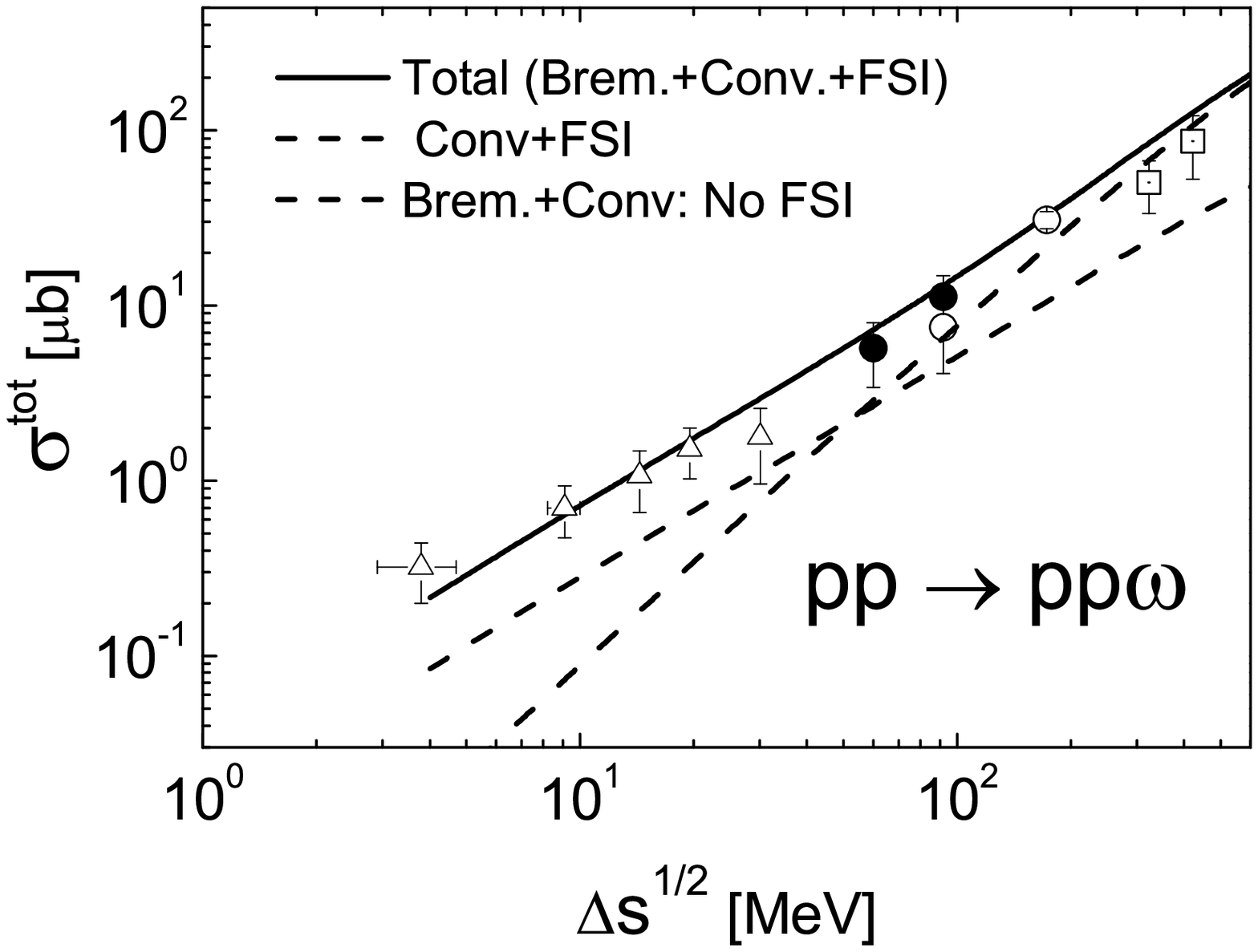} %
\includegraphics[width=.5\textwidth]{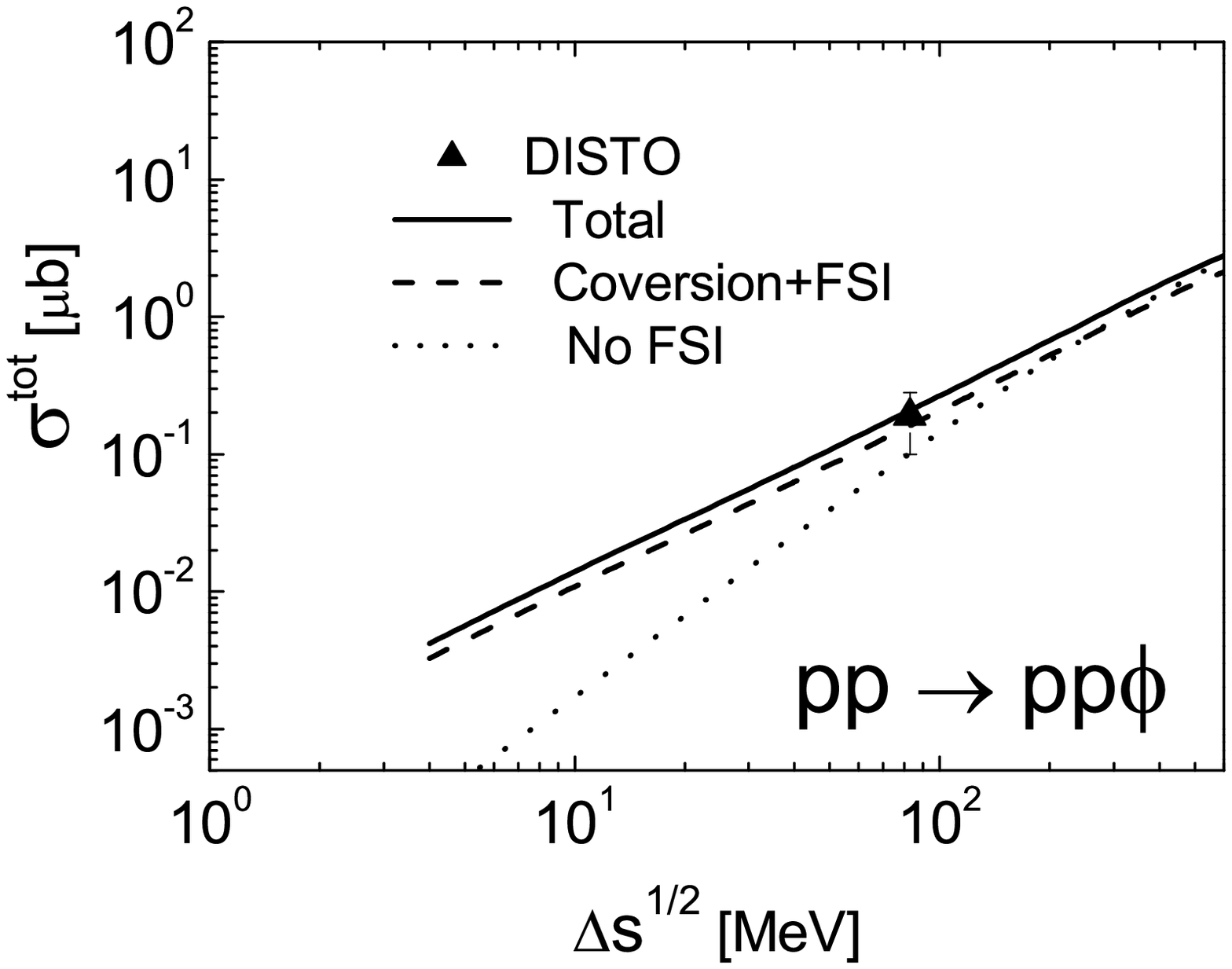} %
\includegraphics[width=.5\textwidth]{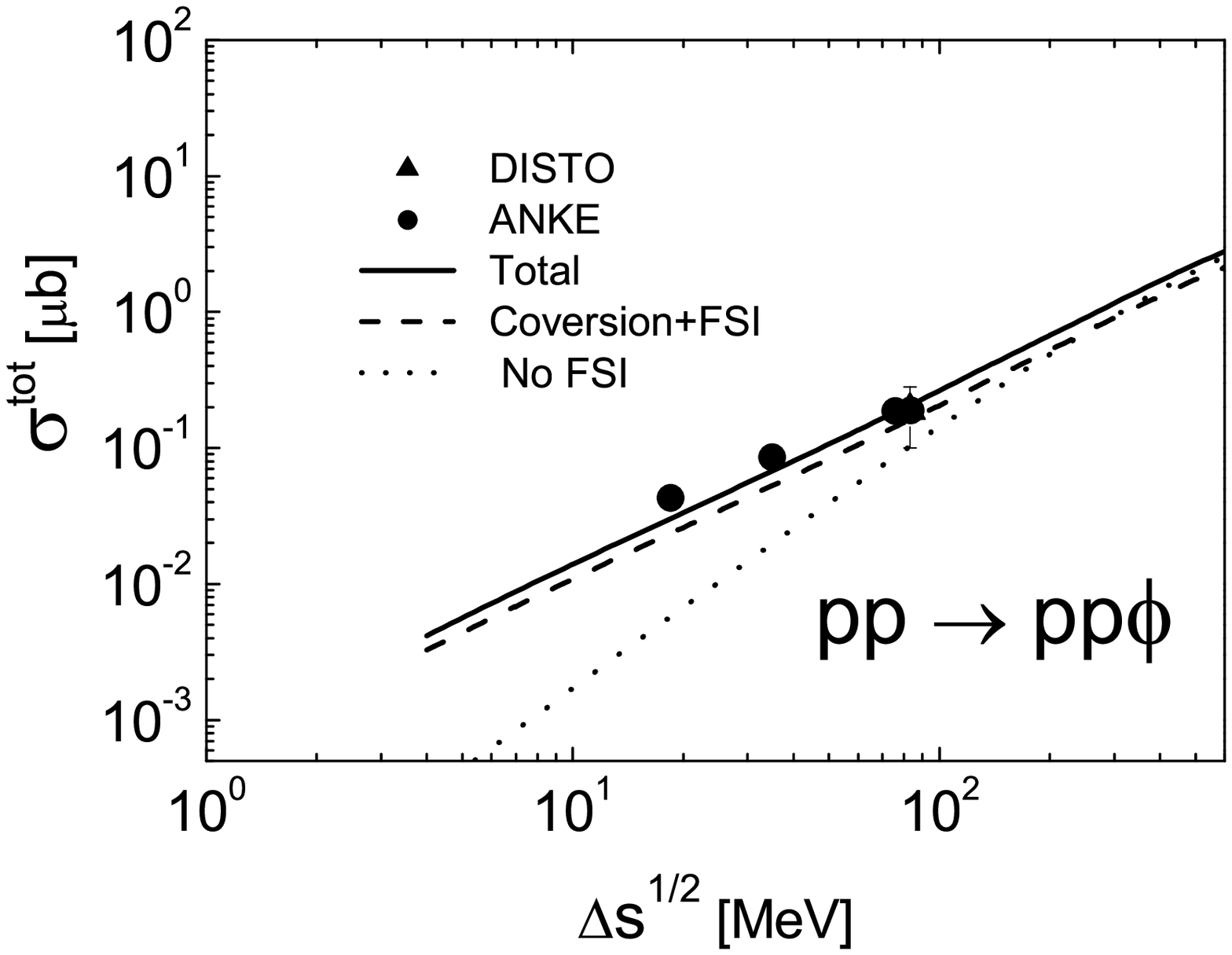} %
\vskip -1mm
\caption{Cross sections for $\omega$ (top) and $\phi$
(bottom) production from \cite{our_omega,our_omega_phi} (left,
data for $\omega$ from
\cite{TOF} (open circles), \cite{hibou} (triangles) and \cite{disto1} (squares)
and for $\phi$  from \cite{disto1,disto}).
The new data situation confirms these predictions (right,
with data for $\omega$ from \cite{ankenew}
and for $\phi$ from \cite{newphi}, both ones depicted as filled circles).}
\label{OBE}
\end{figure}

\subsection{One-boson model at work}

These seemingly many ingredients (coupling strengths, formfactors
and their cut-offs, see \cite{our_eta,our_eta_prime})
may cause the impression that the
one-boson exchange approach to hadronic observables near threshold
does not have too much predictive power. Two counterexamples may lend more
credibility to the approach. In Fig.~\ref{OBE}
the model results of \cite{our_omega,our_omega_phi} are exhibited (left panels).
Later on the data basis has been improved confirming the model predictions (right panels).
Further applications of the present approach to $\omega$ and $\phi$
production involving a final deuteron and including polarization observables,
have been presented in \cite{our_deuteron_phi}, while
\cite{our_bremsstrahlung} extends the formalism to virtual
bremsstrahlung in $NN \to NN \gamma* \to NN e^+ e^-$ reactions.

\begin{figure}[h]  
\vskip -3mm
\includegraphics[width=.53\textwidth]{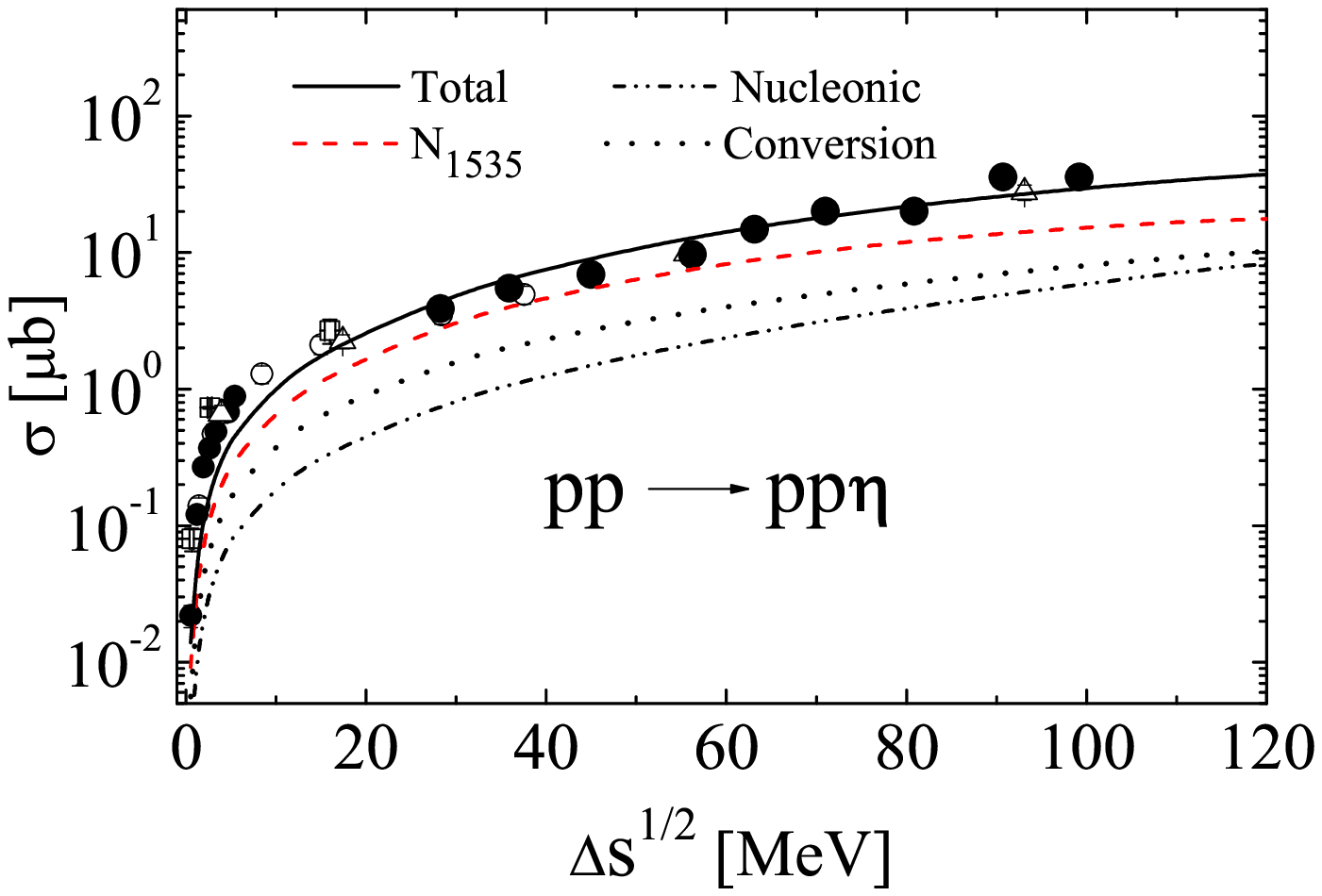} %
\hspace*{-4mm}
\includegraphics[width=.53\textwidth]{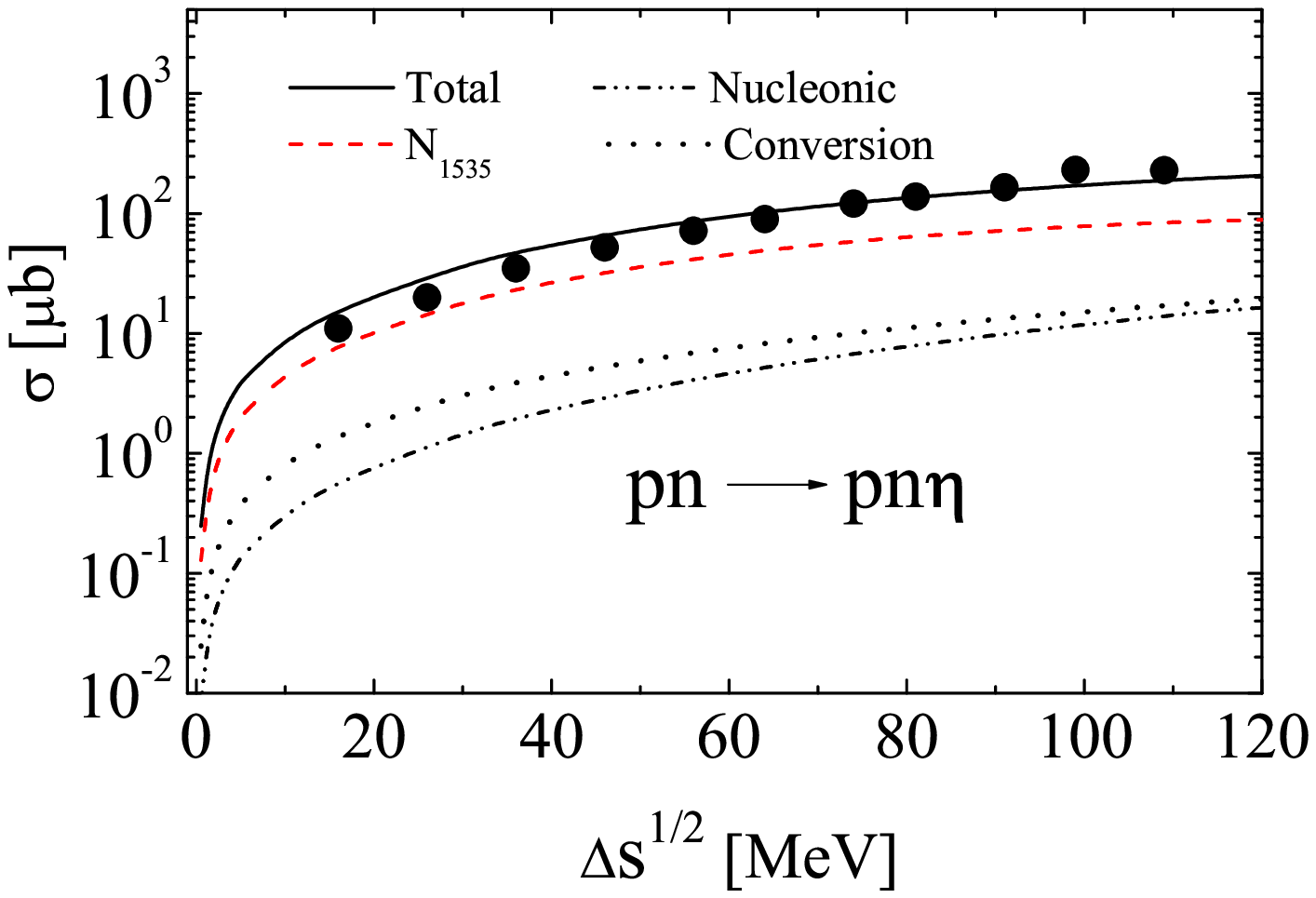} \\ %
\vskip -6mm
\includegraphics[width=.53\textwidth]{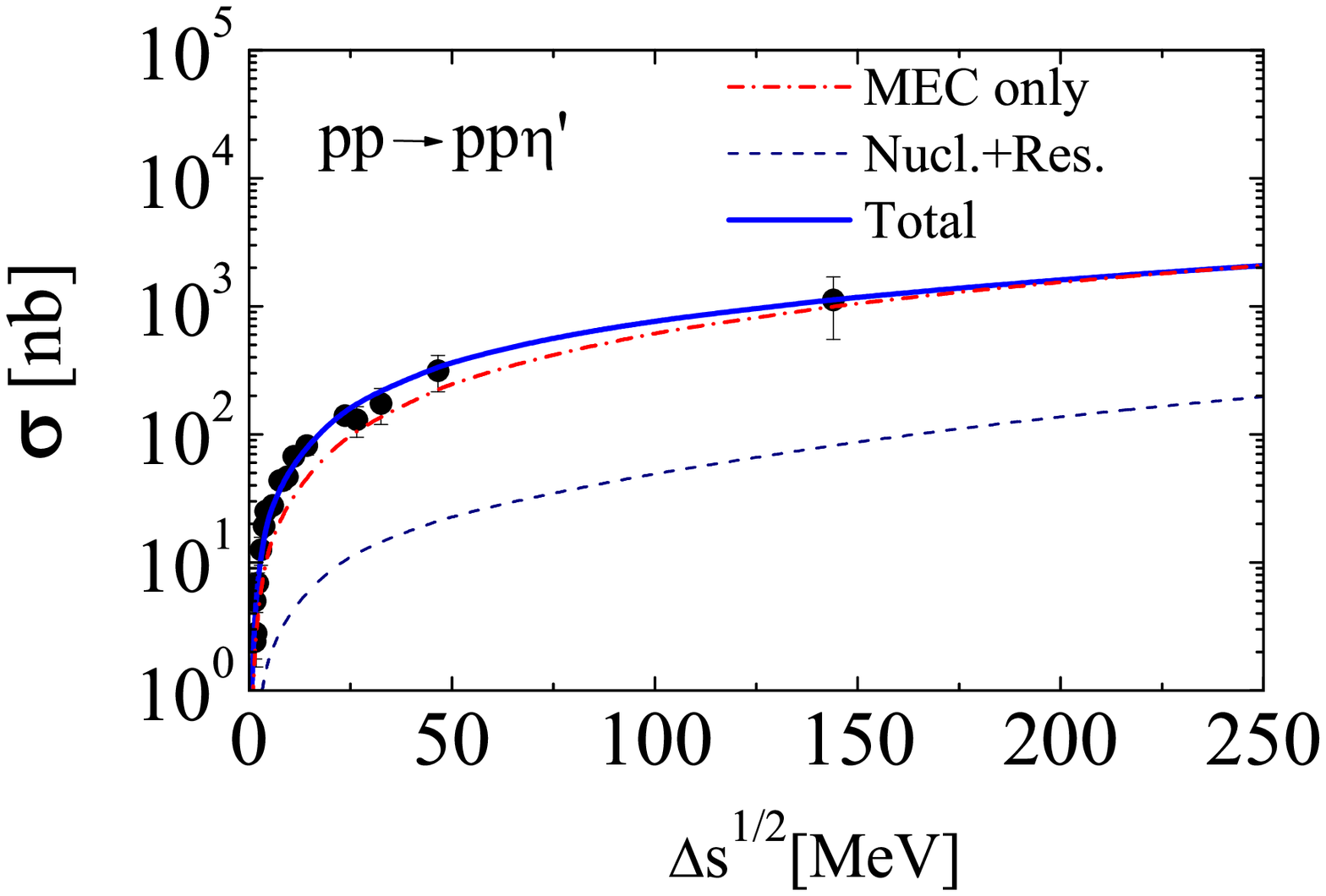} %
\hspace*{-4mm}
\includegraphics[width=.53\textwidth]{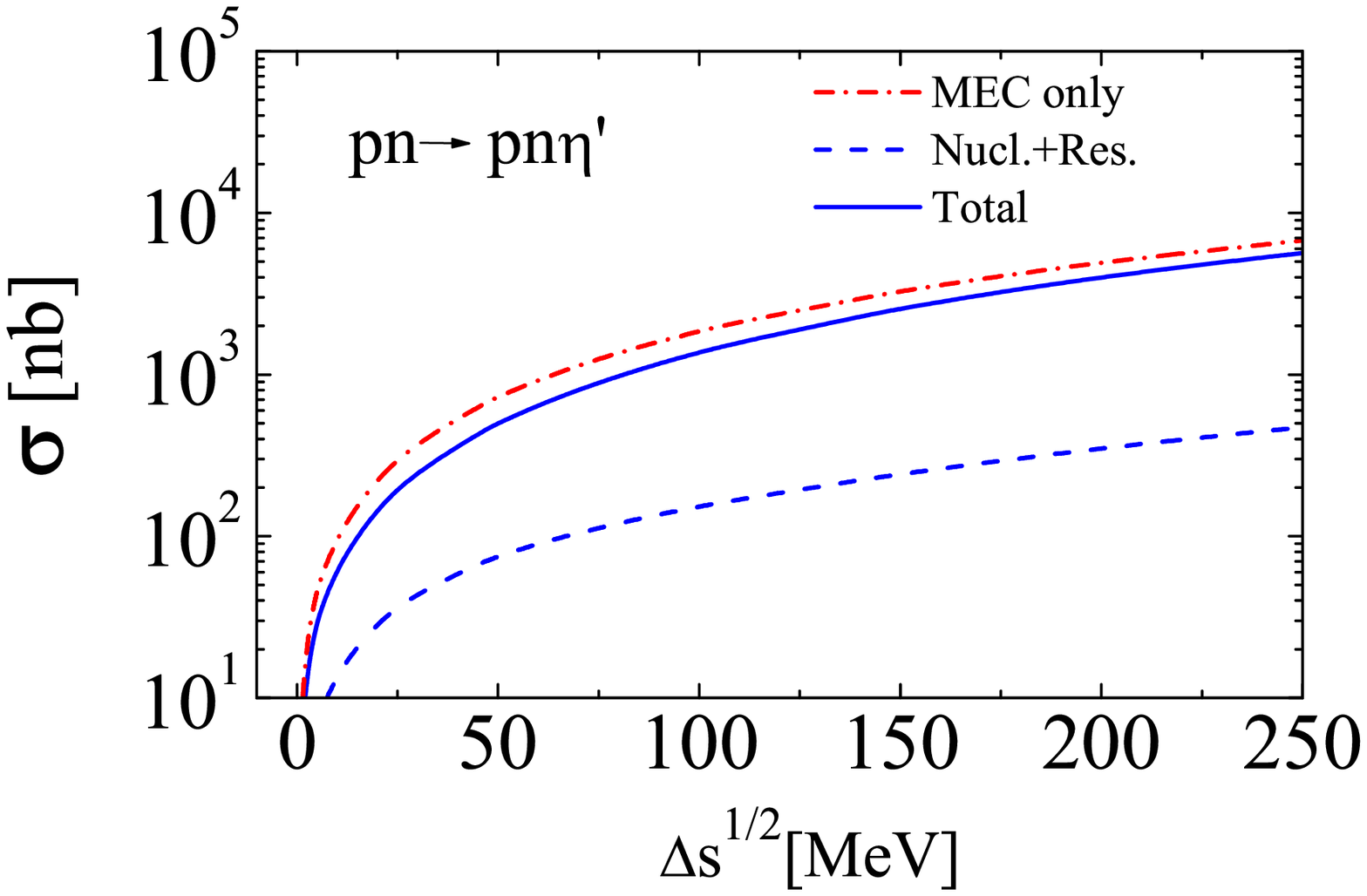} %
\vskip -3mm
\caption{Total cross sections for $\eta$ (top) and
$\eta$' (bottom) production
as a function of the energy excess
in $p+p$ (left) and $n + p$ reactions (right).
For data quotation and further details cf.\ \cite{our_eta,our_eta_prime}.}
\label{fig1_-2_}
\end{figure}

\section{Results}

Numerical evaluation of the given formalism results in the total
cross sections exhibited in Fig.~\ref{fig1_-2_}.
Available data 
are nicely reproduced in the $p + p$ channel
(a concern could be the region of excess energy $\Delta s^{1/2} \sim 10$ MeV for $\eta$).
Since now new parameters enter, the channel $n + p$ represents a prediction,
in agreement with data in case of $\eta$; data are not (yet) available for $\eta$'.

\begin{figure}[h]  
\vskip -6mm
\includegraphics[width=.5\textwidth]{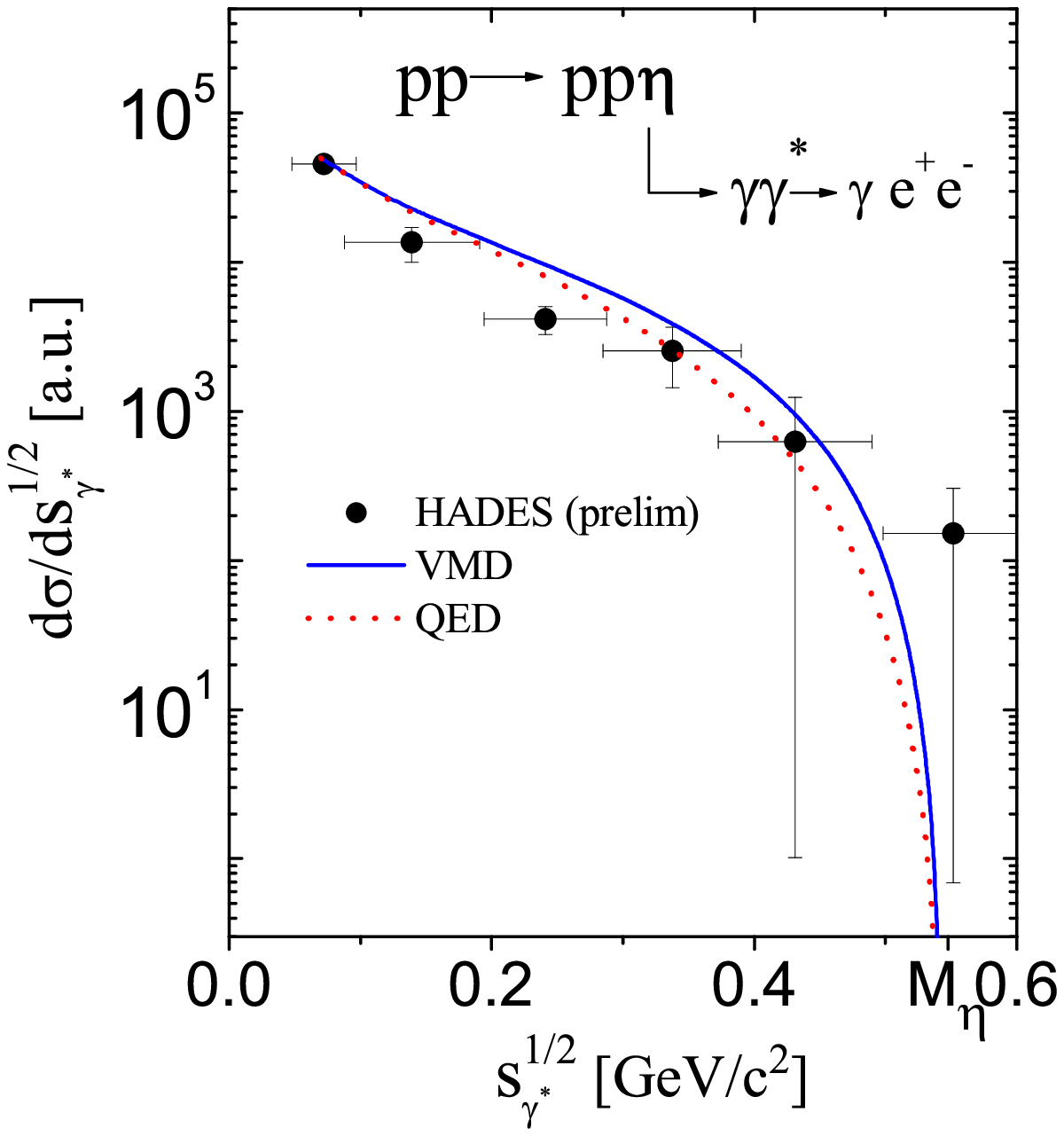} %
\includegraphics[width=.53\textwidth]{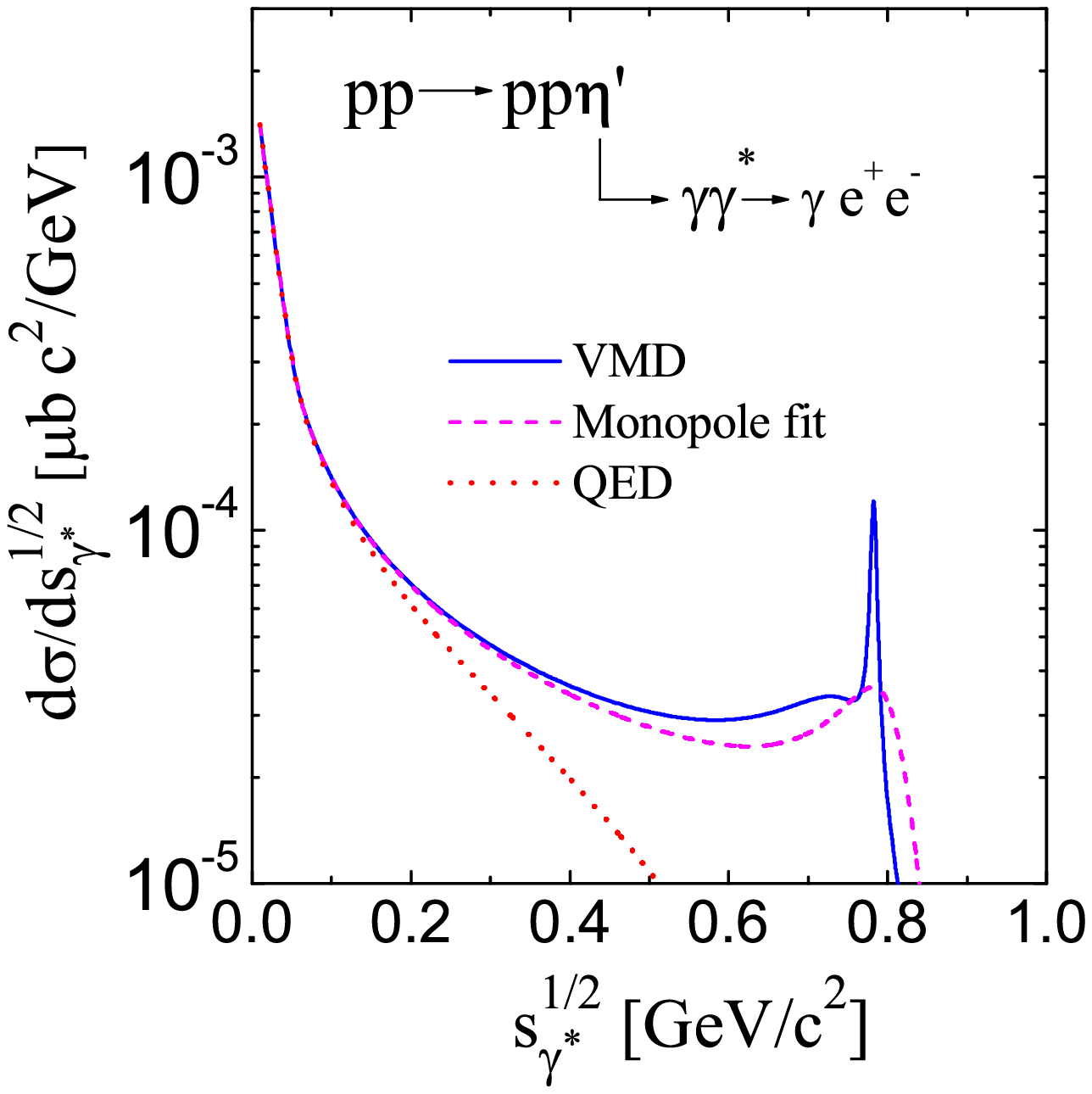} %
\vskip -3mm
\caption{Differential cross sections for $\eta$ (left, HADES data from \cite{Ingo},
for $T_p = 2.2$ GeV)
and $\eta$' (right, for $T_p = 2.5$ GeV)
which give access to the formfactors. }
\label{fig7-8_}
\end{figure}
The cross sections $d \sigma / d s^{1/2}_{\gamma*}$, resulting from
the integration of (\ref{two}) over $s_{ps}$, are exhibited in Fig.~\ref{fig7-8_}.
There is a tiny difference when neglecting the internal strong interaction
structure of $\eta$ (''QED'' formfactor) or when using the ''VMD'' formfactor,
see left panel. The situation changes drastically for $\eta$'. Here,
the account of the internal structure becomes important, see right panel.
Precision data would even allow for a sensible test of the VMD hypothesis.
It has been shown in \cite{our_eta,our_eta_prime} that the formfactors
can be deduced from given cross sections $d \sigma / d s_{\gamma*}^{1/2}$.

\section{Summary}

In summary we report on calculations of the reaction $NN \to NN ps$
with $ps = \eta, \eta'$ and subsequent Dalitz decay $ps \to \gamma e^+ e^-$
within a one-boson exchange model. We point out that isolating $\eta$
and $\eta$' contributions, e.g., in $p + p$ collisions, allows for an
experimental determination of the transition formfactors
$F_{ps \gamma \gamma*}$. In particular, for $\eta$' the vector meson dominance
hypothesis would be testable. On the other hand, the $\eta$ Dalitz decay
channel is a strong source of $e^+ e^-$ pairs in medium-energy heavy-ion
collisions which need to be understood before firm conclusions on possible
in-medium modifications of hadrons can be made.
We emphasize that, once the model parameters are adjusted in the $p+p$ channel,
the $n+p$ channel is accessible without further parameters.

For further improvements of the presented formalism we refer the interested reader
to \cite{Nakayama}, where $N + N$ collisions and $\eta, \eta'$ photo-production
are considered on a common footing.

\end{document}